\documentclass[twocolumn,amsmath,showpacs,amsfonts,aps,prc,floatfix]{revtex4}
\usepackage{graphicx}
\usepackage{bm}

\begin{document}

\title{The phase diagram of QCD in  a BAG+HRG based equation of state: appearance of a pseudo-critical point}
 \author{Partha Pratim Bhaduri}
\email[E-mail:]{partha.bhaduri@veccal.ernet.in} 
\author{Victor Roy}
\email[E-mail:]{victor@veccal.ernet.in} 
\author{A. K. Chaudhuri}
\email[E-mail:]{akc@veccal.ernet.in}

\affiliation{Variable Energy Cyclotron Centre, 1/AF, Bidhan Nagar, 
Kolkata 700~064, India}

\begin{abstract}

Mapping the QCD phase boundary and locating critical end point still remains as an open problem in strong interaction physics. Predictions about the co-ordinates of the critical point in the $(T, \mu_B)$ plane, from different QCD motivated models show a wide variation. Lattice QCD calculations are also available, that give an estimation of the critical point for chiral phase transition, where the transition changes its nature from rapid cross over to first order transition. Recently co-ordinates of the critical point for deconfinement phase transition are claimed to be found as an endpoint of the first order phase transition line, in Bag model scenario. In the present paper we have shown that Bag model gives a complete first order phase transition line in the $(T, \mu_B)$ plane, and one can not have any point where the transition changes its nature. 
\end{abstract}

\pacs{47.75.+f, 25.75.-q, 25.75.Ld} 

\date{\today}  

\maketitle

In recent years there is much interest in QCD phase diagram and in particular about the location of QCD critical end point. That a critical end point exists in the QCD phase diagram can be argued as follows: finite temperature lattice simulation of baryon free ($\mu_B=0$) QCD indicate a cross-over transition at $T_c\approx 170\pm 20$ MeV
\cite{Cheng:2007jq}, \cite{Soltz:2009bc}, \cite{Aoki:2009sc}, \cite{Aoki:2006br}. On the other hand there are indications that at zero temperature but at finite baryon density, QCD has a 1st order transition \cite{Asakawa:1989bq}, \cite{Berges:1998rc}, \cite{Barducci:1989eu}. Only with the existence of a critical end point where the 1st order transition ends, the cross over transition at $(T,\mu_B=0)$ can be reconciled with a 1st order transition at $(T=0,\mu_B)$. Unfortunately, lattice simulations at finite baryon density is plugged with the well known fermion sign problem and at the present state of art, location of QCD phase boundary or the critical end point is beyond the lattice simulations. Several methods have been devised to surpass the fermion problem, e.g Taylor expansion \cite{Gavai:2004sd}, \cite{Gavai:2003mf}, re-weighting \cite{Barbour:1997bh}, \cite{Allton:2002zi}, \cite{Fodor:2001pe}, imaginary chemical potential \cite{de Forcrand:2002ci}, \cite{de Forcrand:2003hx}, \cite{deForcrand:2006pv}. However, it is uncertain whether the methods mentioned above are valid at large baryon density. QCD phase diagram has been studied in QCD inspired models, e.g. instanton models, linear sigma model, Nambu-Jona-Lasinio (NJL) or Polyakov loop extended Nambu-Jona-Lasinio models  \cite{Berges:1998rc}, \cite{Alford:1997zt}, \cite{Rapp:1997zu}, \cite{Asakawa:1989bq}, \cite{Meisinger:1995ih}, \cite{Ghosh:2006qh}. In QCD inspired model, in general quark degrees of freedom are integrated out. The model results depend on the cut off parameter. Depending on the model parameters, QCD phase diagram and location of critical end point vary widely. 

In a recent work, Singh et. al. \cite{Singh:2009jd}, have obtained the QCD phase boundary for de-confinement transition by using Gibb's equilibrium criterion for first order phase transition between QGP and hadronic phase. The hadronic phase is modeled by a gas of interacting hadrons, where the geometrical size of the baryons is explicitly incorporated as the excluded volume correction, in a thermodynamically consistent manner. All baryons, mesons and their resonances having masses up to 2 GeV are included in calculation and strangeness conservation is taken into account by equating the net strangeness to zero. Mesons are considered as point-like particles, where as an equal volume  $v_{ex}=\frac{4\pi}{3}r_c^3$  is assumed for each type of baryon with a hard-core radius $r_c = 0.6 - 0.8 fm$. In addition, full quantum statistics is included in the partition function of the grand canonical ensemble which helps to navigate the region of the phase diagram with low T and high $\mu_B$. Thermodynamic functions of weakly interacting quark matter is obtained by a simple Bag model equation of state (EOS) with perturbative corrections of the order of $\alpha_{s}^{3/2}$ in strong interaction coupling constant $\alpha_{s}$. Having modeled the two phases, the QCD phase boundary is obtained advocating Gibb's equilibrium criterion for first order phase transition. Gibb's criterion demands that at the transition point the pressure of the hadronic state $P_H(T,\mu_B)$ and that of the qgp state $P_Q(T,\mu_B)$ becomes equal. 

Starting from a low but non-zero value of T and large value of $\mu_B$, gradually they have moved towards large T and small $\mu_B$. At a given T, a corresponding value of $\mu_B$ is found out at which the pressure equality holds and beyond which QGP pressure dominates. It has been observed that the line of co-existence between two phases ends at a point which is being interpreted as the end point of the 1st order transition line, or the critical end point. Thus the precise co-ordinates of the QCD critical point has been claimed to be estimated for deconfinement phase transition in the $\mu_B-T$ plane. The critical values of temperature and chemical potential are found to be $(T_c = 160 MeV, \mu_c = 156 MeV)$ for hard-core radius $r_c = 0.6 fm$ and Bag constant $B^{1/4} = 216 ~MeV$. For  $r_c = 0.8 fm$ and $B^{1/4} = 200 ~MeV$, the point is shifted down in temperature to $(T_c = 146 MeV, \mu_c = 156 MeV)$. Variation in QCD scale parameter$\Lambda$ is reported to give insignificant change in the location of the phase boundary. Considering that earlier investigation \cite{Satz}, \cite{Dixit}, \cite{Cleymans1}, \cite{Magas}, \cite{Castorina} of phase diagram in terms of Bag model EOS and hadronic resonance gas fails to detect the QCD critical point, we have studied the model in detail. We find  that the simple interpretation of minimum of the chemical potential below which co-existence line ceases to exist as the critical end point is misleading and appears due to lack of sufficient numerical precision.


In the present work we have employed the same Bag model EOS as used in \cite{Singh:2009jd} to compute the thermodynamic parameters of the partonic phase. In this model, QGP is assumed to consist of massless quarks (u,d), their antiquarks and gluons only and the pressure in the QGP phase takes the form

\begin{equation}
  \begin{split}
 P_{QGP}
= \frac{37}{90}\pi^2 T^4 + \frac1{9}\mu_B^2 T^2 + \frac{\mu_B^4}{162 \pi^2} 
\\
- \alpha_S\left[\frac{11}{9}\pi T^4 + \frac{2}{9\pi^2}\mu_B^2 T^2 + \frac1{81\pi^3}\mu_B^4\right ]
\\
+\frac{8\alpha_S^{3/2} T}{3\pi^2\sqrt{2\pi}}{\left[\frac{8\pi^2 T^2}{3} +\frac2{9}\mu_B^2\right ]^{3/2}} - B
 \end{split}
\end{equation}
where $\mu_B$, T dependence of $\alpha_S$ can be given as [15]:

\begin{equation}
\alpha_S
= \frac{12\pi}{29}\left[ln  (\frac{0.089 \mu_B^2 + 15.622 T^2}{\Lambda^2})\right]^{-1}
\end{equation}
\\
Here we have used $B^{1/4} = 200 ~MeV$ and $\Lambda = 100 ~MeV$ in our calculation.

For simulating the hadronic phase we considered a hadron resonance gas model (HRG) which includes all hadrons and their resonances having masses up to 2.5 GeV. Excluded volume correction has also been incorporated following the prescription by Cleymans and Suhonen \cite{Cleymans2}. In this approach, for a given eigen volume $v_{ex}$, the excluded volume corrected pressure, is obtained as,

\begin{equation}
P^{excl}_H(T,\mu_B)=\frac{P^{id}_B(T,\mu_B)}{1+v_{ex} n_B^{id}(T,\mu_B)} + P^{id}_M(T,\mu_B) \\
\end{equation}

Here $P^{id}_{B(M)} (T,\mu_B)$ is the pressure of the corresponding the ideal gas having point like baryons (mesons) and can be computed as

\begin{eqnarray}
P^{id}_{B(M)}(T,\mu_B)&=&
\sum_i\frac{g_i}{(2\pi)^3}\int d^3p \frac{p^2}{3\sqrt{m_i^2+p^2}} \nonumber\\
&\times&\frac{1}{e^{(\sqrt{m_i^2+p^2}-\mu_i n_i)/T}\pm 1}
\end{eqnarray}

where the sum is over all the baryonic (mesonic) species and their resonances included in  the calculation.

Though not thermodynamically consistent, but the resulting reduced pressure obtained by this approach is in close agreement with the approach employed by Singh et al \cite{Madhukar}, which preserves thermodynamical consistency. Having computed the pressure of the two phases, we then obtain the phase diagram, in $\mu_B-T$ plane, by searching for zeros of the equation, 

\begin{equation} \label{eq5}
\Delta P(T,\mu_B) = P_{QGP} (T,\mu_B)-P^{excl}_H (T,\mu_B)=0
\end{equation}

 The locus of the zeros is identified as the phase boundary in $\mu_B-T$ plane along which QGP and the hadronic phase can co-exist. The numerical zero-search can be employed in two ways. We can fix a temperature $T_0$ and vary $\mu_B$ over the specified range in steps of $\delta\mu_B$ and calculate $\Delta P(T_0,\mu_B)$ and $\Delta P(T_0,\mu_B+\delta\mu_B)$. If they are of opposite sign then there must a value of $\mu_B=\mu_B^0$, such that $\Delta P(T_0,\mu_B^0)=0$, and which can be calculated using straight line interpolation. Then ($T_0,\mu_B^0$) denotes a first order phase transition point in the $\mu_B-T$ plane, provided $\Delta P(T_0,\mu_B+\delta\mu_B) > \Delta P(T_0,\mu_B)$. We can then go to another temperature $T_0+\delta T$ and get the corresponding value of $\mu_B^0$. In this way we can trace the first order phase transition line in $\mu_B-T$ plane. Let us name this method as $\mu_B$-scan. This is identical to the method employed by Singh et. al. to construct the QCD phase digram. On the contrary one can in principle fix a baryon chemical potential $\mu_B^0$ and scan the temperature axis identically to obtain the corresponding transition temperatures. We can call this approach as T-scan.

\begin{figure}[t]
\center
\resizebox{0.5\textwidth}{!}{%
  \includegraphics{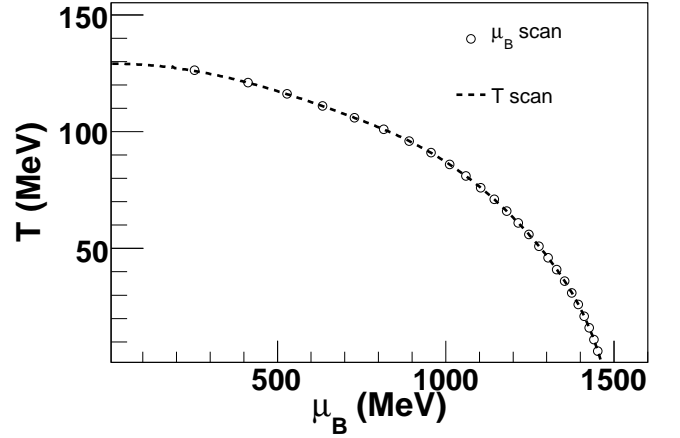}
}
\caption{(color online)
\label{fig1} The estimated QCD phase boundary for first order deconfinement phase transition, with $B^{1/4} = 200 ~MeV$ and $v_{ex} = 1 ~fm$,  obtained through both T-scan and $\mu_B$-scan. The step length for search is kept same for both cases (see text).} 
\end{figure}

 We have employed both methods namely T-scan and $\mu_B$-scan to draw the phase diagram in the  $\mu_B-T$ plane. For both cases we have fixed the step length of scan as $\delta T = \delta\mu_B = 5$ MeV. The resulting phase boundaries are shown in Fig.~\ref {fig1}.

\begin{figure*}
  \centering
  \includegraphics[height=6.5cm, width=6.5cm]{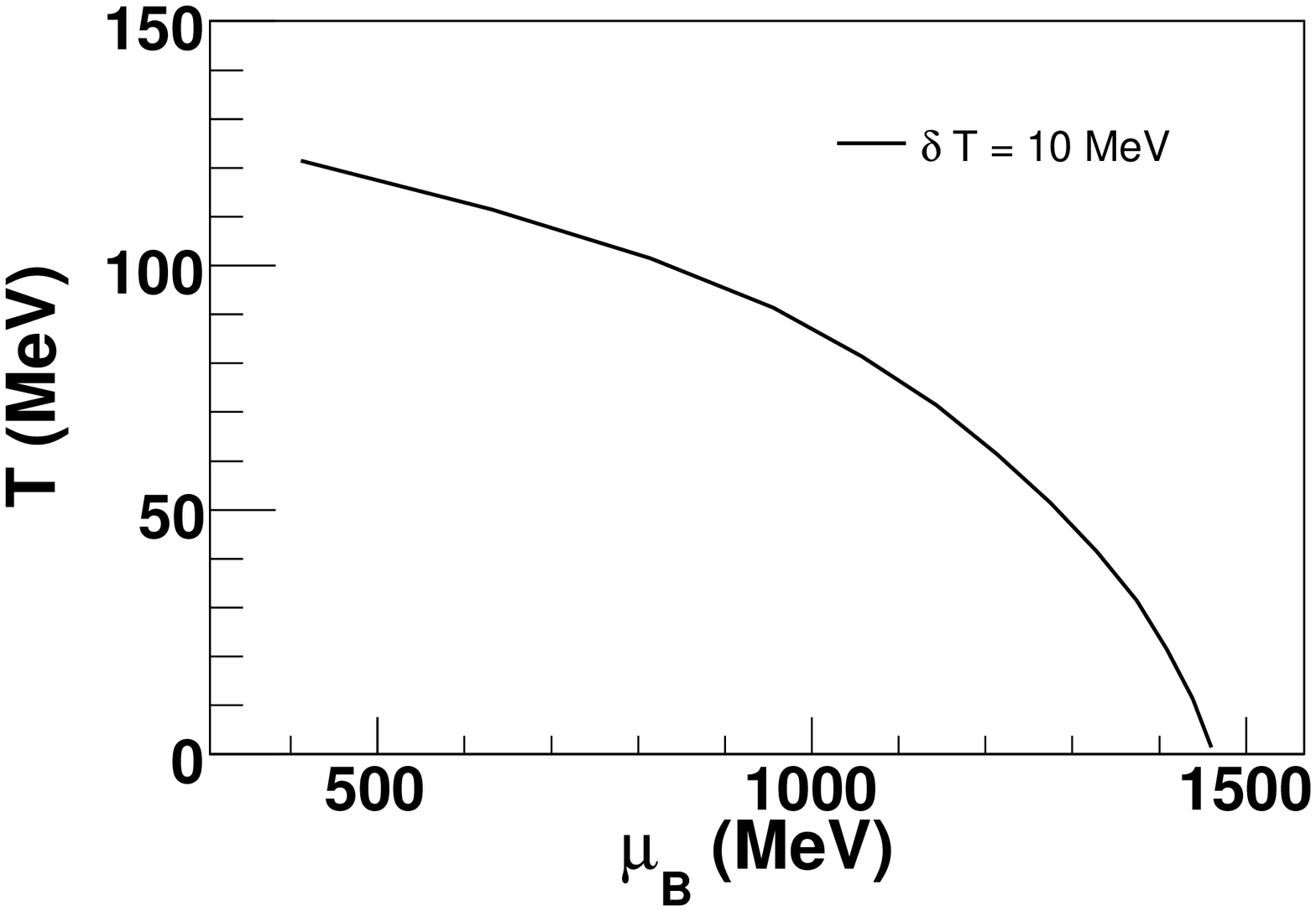}
  \label{fig:F2a}
  \includegraphics[height=6.5cm, width=6.5cm]{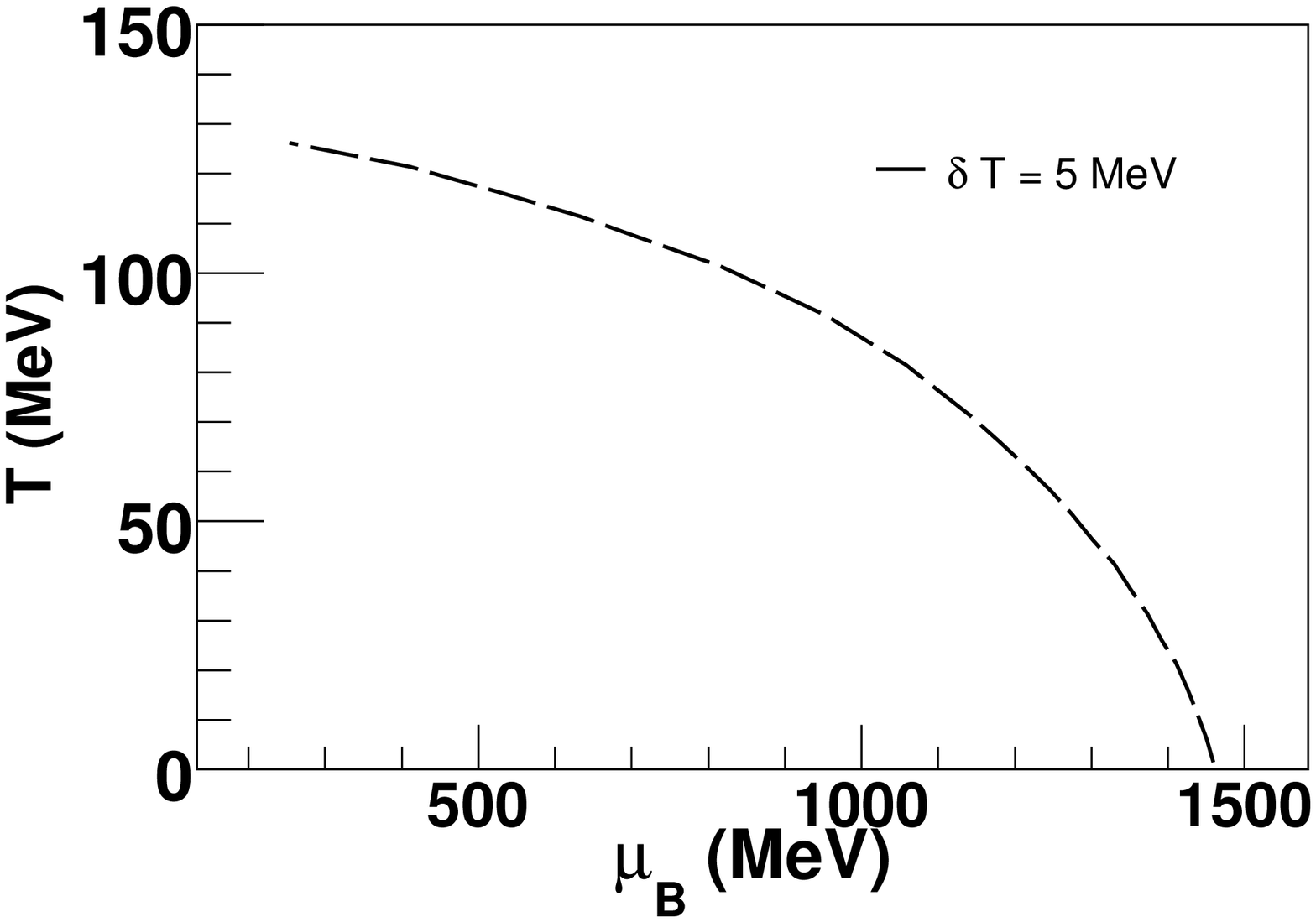}
  \label{fig:F2b}
  \includegraphics[height=6.5cm, width=6.5cm]{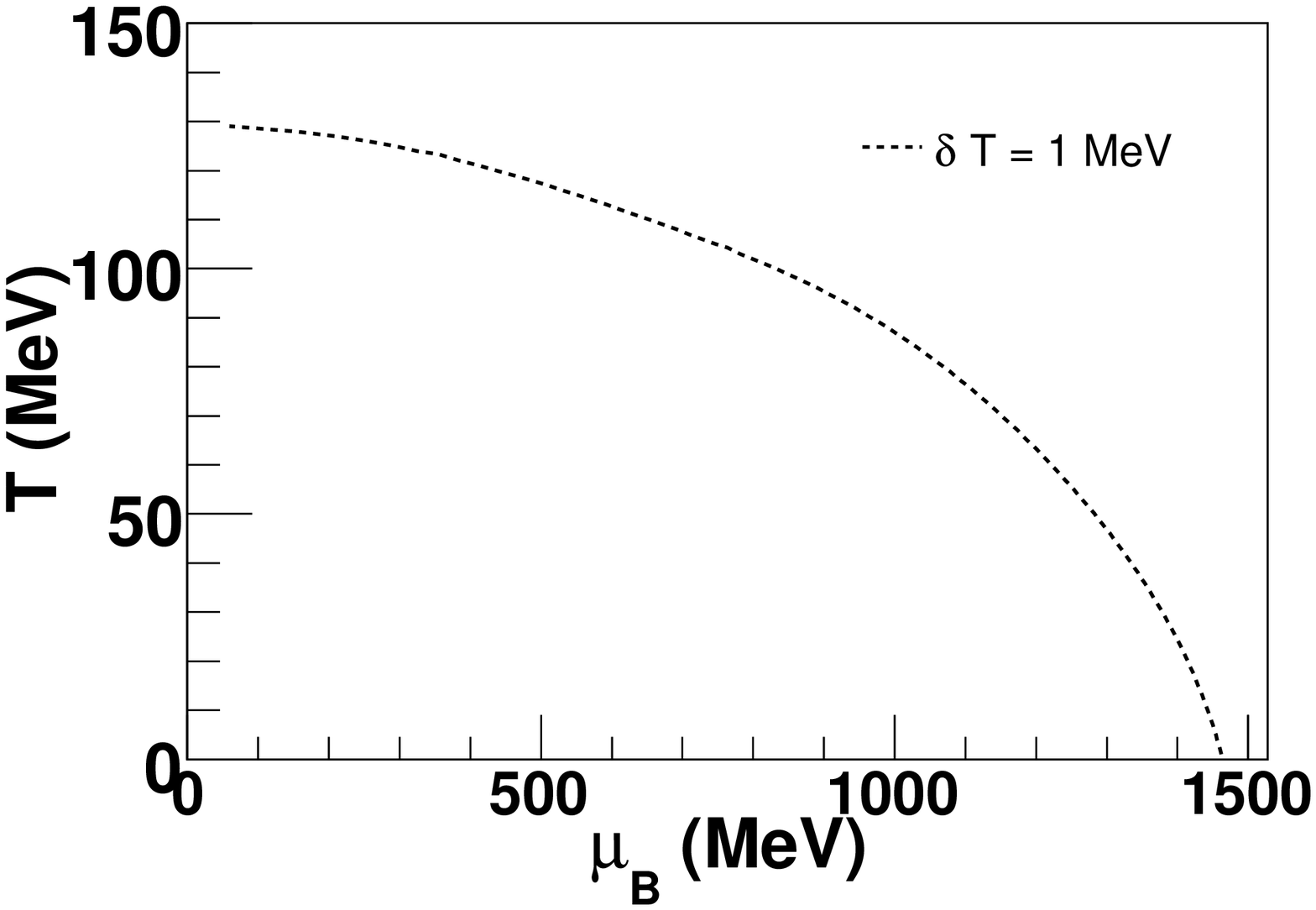}
  \label{fig:F2c}
  \includegraphics[height=6.5cm, width=6.5cm]{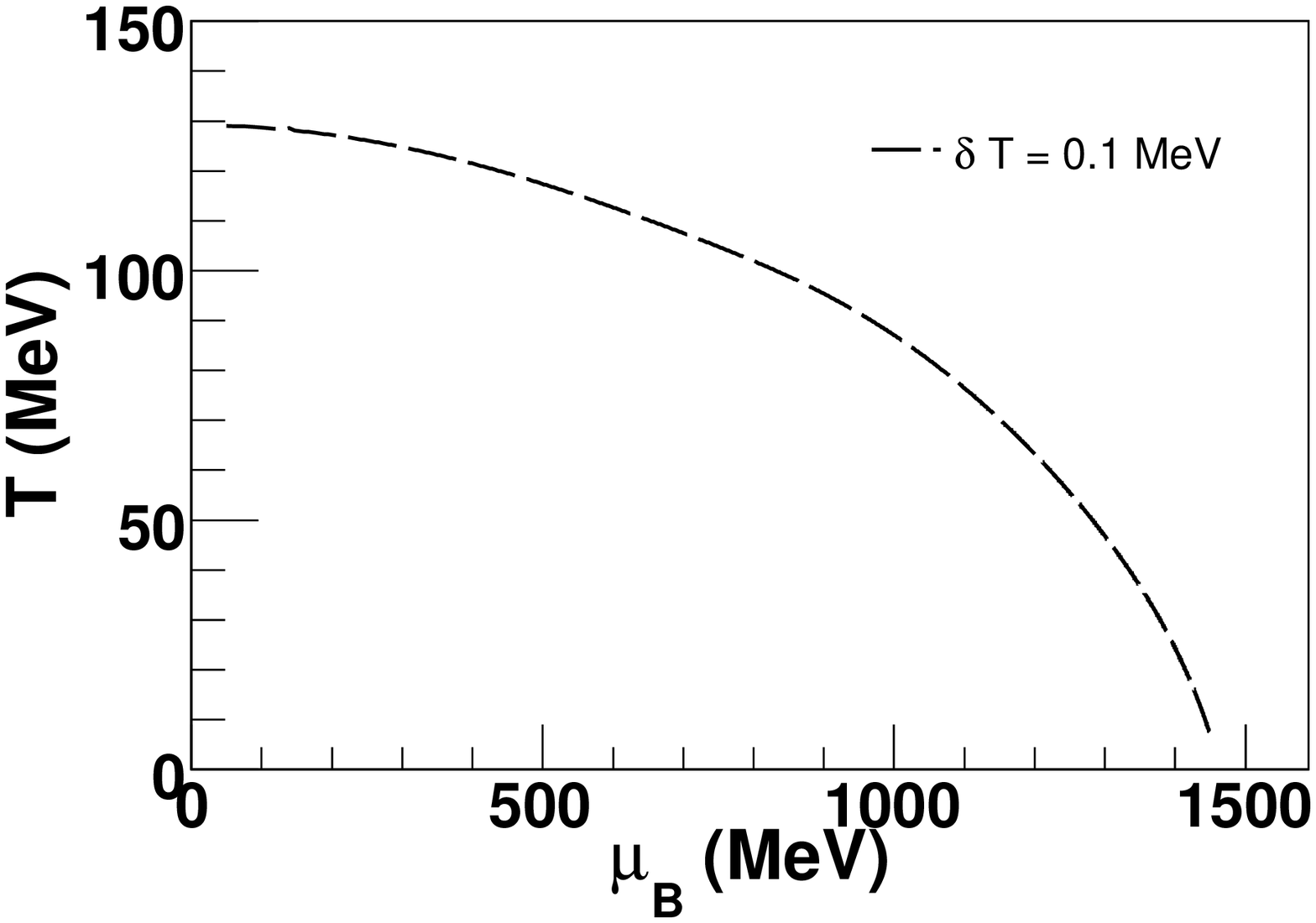}
  \label{fig:F2d}
  \caption{\label{fig2}The QCD phase boundary obtained from $\mu_B$-scan, for gradually decreasing values of $\delta T$. As $\delta T$ decreases, the line gradually approaches towards temperature axis. For all the cases we have assumed $B^{1/4} = 200 ~MeV$ and $v_{ex} = 1 ~fm$.}
\end{figure*}

The first order phase transition line obtained from $\mu_B$ scan ends at some non-zero value of $\mu_B$, the so called critical point of the de-confinement phase transition. However T-scan with same step length yields a phase boundary that reaches up to $\mu_B=0$. Hence the end point of the first order phase transition line as obtained through $\mu_B$ scan can no longer be interpreted as the true critical point of de-confinement phase transition. Rather it is a pseudo-critical point, which appears due to insufficient numerical resolution of the method employed for the construction of the phase diagram. The absence of 1st order phase transition points (zeros of $\Delta P (T, \mu_B)$) beyond some minimum of  $\mu_B$ can be understood by looking at the structure of the phase diagram obtained through T-scan. As the line approaches towards the T axis it becomes more and more flat in T. Hence the zeros of $\Delta P$ becomes more and more closer in T and lie with in the search length for a step length of $\delta T=5$ Mev. If one gradually decreases the step length, more and more zeros start to appear in the low $\mu_B$ region and corresponding curve eventually moves towards the $\mu_B = 0$ axis. This becomes evident from Fig.~\ref{fig2}, where we have plotted the phase transition line for gradually decreasing values of $\delta T$. Thus the appearance of the end point of the first order phase transition line in $\mu_B-T$ plane, is not physical. This is rather a pseudo-critical point, coming into the picture due to lack of numerical detectability.




\begin{figure}[t]
\vspace{0.4cm} 
\resizebox{0.5\textwidth}{!}{%
  \includegraphics{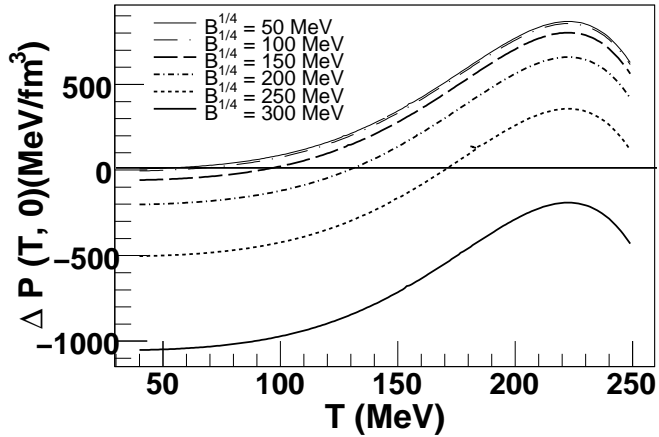}}
\caption{\label{fig3} Variation of $\Delta P$ with temperature at zero chemical potential for different values of Bag constant.}
\end{figure}
\vspace{0.4cm} 

Indeed, one can argue that for bag model type equation of state, it is possible to tune the bag pressure such that Eq.\ref{eq5} is satisfed for zero baryons density.  In Fig.~\ref{fig3}, for $B^{1/4}$=(50-300) MeV, we have plotted the pressure difference $\Delta P(T) = P_{QGP}(T,\mu_B=0) - P_{H}(T,\mu_B=0)$ as a fuction of temeprature. One observes that for $50 MeV \leq B^{1/4} \leq 270$ MeV, a co-existence phase exists. At low temperatures, QGP pressure is smaller than the hadron pressure and $\Delta P(T)$ is negative. At high temperature QGP pressure dominates over the hadronic pressure resulting a positive $\Delta P(T)$. However, the transition temperaure shifts to higher $T$ with increasing $B$ and there by indicating a Bag pressure dependent first order de-confinement phase transition at zero baryon chemical potential. At sufficiently high (low) values of B $(B^{1/4}= 300 (50) MeV)$, the QGP pressure is always lower (higher) than the hadronic pressure (at $\mu_B = 0$), and one can not get any first order phase transition. But over a reasonable range of Bag constants, one can always get a solution of $\Delta P(T,0) = 0$ and hence a first order hadron to QGP phase transition at $\mu_B = 0$. At very high values of temperature $\Delta P(T)$ is seen to decrease with T. This can be attributed to the larger degrees of freedom associated with the hadronic phase compared to the QGP because of the exponential growth of the hadrons and resonances at very high T, resulting in a higher pressure in HG than QGP.This indicates a reversal phase transition from QGP to hadron gas at a still higher temperature \cite{Madhukar}. But once the system goes over to the QGP phase, due to rise in temperature, it should continue to stay in that phase owing to the asymptotic freedom of QCD. Moreover, it is expected that the hadronic interactions are become significant when hadrons are closely packed in a hot and dense hadron gas. This anomalous behavior arises owing to the treatment of the hadronic phase as an ideal gas of non-interacting point-like hadrons. As a result of this assumption, the thermal production of an arbitrarily large number of hadrons in a given volume at very high T (or $mu_B$) is possible and eventually leads to infinitely large energy densities and pressure. In fact a simple remedy to this problem is the inclusion of finite, proper volume for each hadron, which leads to a hard-core repulsion among themselves at very high temperature and/or density and thereby limiting the number of hadrons in the system so that its volume is completely filled with particles. This is the so called excluded excluded volume correction, where the repulsive force is being incorporated by assigning a geometrical hard-core volume to each hadron. This finally leads to the reduction of the effective pressure and energy density of the hadron gas particularly at high temperature and/or density. We do have incorporated the excluded volume effect in our calculation following the model by Cleymans and Suhonen \cite{Cleymans2}. In this model the the volume correction term is proportional to the net baryon density and thus becomes in-effective at $mu_B = 0$. Hence the hadron gas behaves like an ideal gas and shows the indications of the reverse phase transition at very high temperature.

\begin{figure}
\vspace{0.3cm} 
\resizebox{0.5\textwidth}{!}{%
  \includegraphics{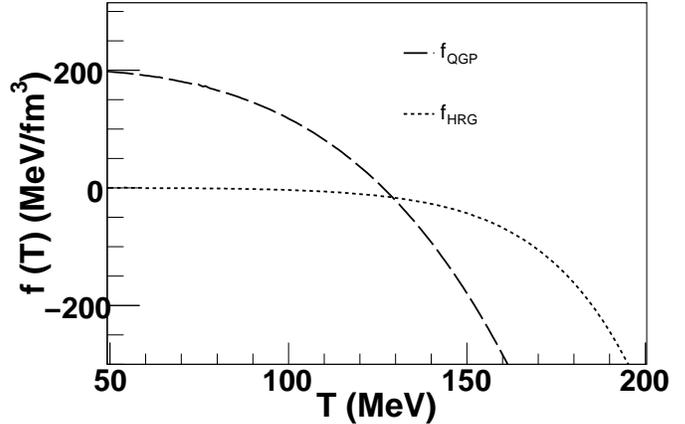}}
\caption{\label{fig4} Variation of the Helmholtz free energy density with temperature at zero chemical potential for both hadronic phase and QGP phase.}
\end{figure}

That our observation of the absence of any true critical point in HRG+Bag model based EOS can be justified further through following arguments. The critical point is the end point of the first order phase transition line and at the critical point transition is believed to be second order. A first order phase transition is always associated with non-zero finite latent heat, which in fact measures the discontinuity in entropy at the transition point. For a second order transition, entropy changes smoothly along the transition point and hence latent heat is zero. In Fig.~\ref{fig4}, we have shown the temperature variation of Helmholtz free energy density ($f=\varepsilon-Ts$) at $\mu_B=0$. Both $f_{had}$ and $f_{QGP}$ decreases as temperature increases. Since the system always follows the path of minimum free energy, hence the system will undergo a transition from hadron phase to partonic phase at $T=129.6$ MeV. Since $s=-\frac{df}{dT}$, hence a kink in f(T) at $T=129.6 MeV$ would indicate a discontinuity in entropy density thereby ensuring the zero chemical potential, temperature driven transition, to be of first order.

To conclude, we would like to clarify that we are not doubting the existence of the critical point in the QCD phase diagram. The ab-initio Lattice QCD calculations have indeed proved that such a point does exist. Though its precise location in the $T-\mu_B$ is a matter of ongoing debate as predictions from different groups vary wildly. Our only motivation is to prove that the explicit construction of the first order phase separation boundary through comparison of a HRG EOS with Bag model EOS, can not give any estimation of the critical end point for de-confinement phase transition.

\end{document}